\documentclass[pra,epsfig,floats,twocolumn,showpacs,amssymb,floatfix]{revtex4}
\usepackage{graphicx}
\usepackage{bm}

\begin{document}

\title{Rectification of the lateral Casimir force in a vibrating
non-contact rack and pinion}

\author{Arash Ashourvan}
\affiliation{Institute for Advanced Studies in Basic Sciences,
Zanjan, 45195-1159, Iran}

\author{MirFaez Miri }
\email{miri@iasbs.ac.ir} \affiliation{Institute for Advanced Studies
in Basic Sciences, Zanjan, 45195-1159, Iran}

\author{Ramin Golestanian}
\email{r.golestanian@sheffield.ac.uk} \affiliation{Department of
Physics and Astronomy, University of Sheffield, Sheffield S3 7RH,
UK}

\date{\today}
\begin{abstract}
The nonlinear dynamics of a cylindrical pinion that is kept at a
distance from a vibrating rack is studied, and it is shown that the
lateral Casimir force between the two corrugated surfaces can be
rectified. The effects of friction and external load are taken into
account and it is shown that the pinion can do work against loads of
up to a critical value, which is set by the amplitude of the lateral
Casimir force. We present a phase diagram for the rectified motion
that could help its experimental investigations, as the system
exhibits a chaotic behavior in a large part of the parameter space.
\end{abstract}


\maketitle


As the technological advances lead to miniaturization of mechanical
devices, engineers face new challenges that are brought about by the
fundamentally different rules that apply at small scales. One of the
biggest problems in small machines is the excessive wear of the many
surfaces that work in contact with each other, which severely
constrains the durability of such machine parts \cite{Wear}. A
particularly attractive idea for overcoming this problem might be to
try and exploit the Casimir effect \cite{Casimir,plu} to transduce
mechanical forces at small scales between different machine parts
that do not have contact with each other. While the classic normal
Casimir force between planar boundaries might not be well suited to
this task, the lateral Casimir force between corrugated
surfaces---which has been predicted \cite{golestan97} and observed
\cite{Mohideen} recently---might be a better candidate. In an
attempt along these lines, we studied the feasibility of a rack and
pinion without contact as a mechanical transducer, and found that
the coupling provided by the lateral Casimir force can make the
pinion stay locked-in with the rack up to surprisingly large
velocities even while doing work against an external load
\cite{AFR1}.

The coupling provided by Casimir force between machine parts is
nonlinear, and could lead to complicated dynamics as already
demonstrated by Capasso and collaborators using the normal Casimir
force \cite{Capasso2}. This implies that any experimental
realization of such force transduction mechanisms should be
carefully guided by theoretical studies of the phase behavior of the
system in the space of tunable parameters. For example, the
nonlinear coupling could very easily lead to chaos, which is
presumably unwanted in such cases and should thus be avoided.

It might also be possible to use the nonlinear properties to our
advantage. A particularly interesting case is the possibility of
producing a net directed output motion (which is non-compact in the
parameter space) by using a periodic (compact) input motion. This
problem is analogous to swimming \cite{childress}, and somewhat
related to the fluctuation--induced ratchets \cite{ratchet}. While
we are primarily interested in directed motion induced by static
Casimir force, we note that dynamic Casimir force
\cite{golestan97,dynCas} can also generate motion by controlled
emission of photons \cite{miri} and getting propulsion from the
back-reaction, as exemplified by the vacuum ``flying carpet'' design
proposed in Ref. \cite{ramin}.

\begin{figure}[b]
\includegraphics[width=0.9\columnwidth]{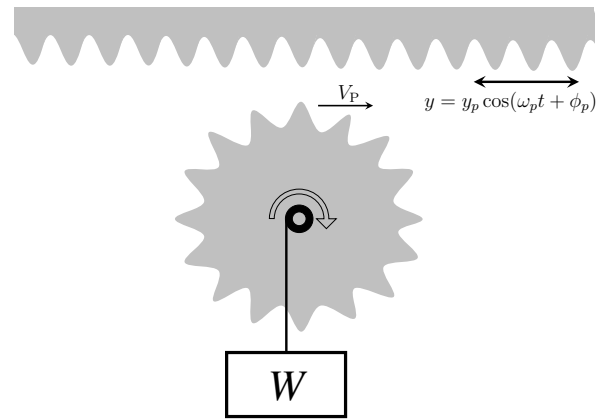}
\caption{The schematics of the non-contact rack and pinion, with the
rack vibrating laterally. Both the rack and the pinion have
sinusoidal corrugations of wavelength $\lambda$. The rectified
motion of the pinion will manifest itself in a positive average
pinion velocity $V_{\rm P}$, while working against an external load
$W$.} \label{fig:gear}
\end{figure}

Here we study the dynamics of a pinion that is kept at a distance
from a vibrating rack, as shown in Fig. \ref{fig:gear}. The two
parts are coupled by the lateral Casimir force between the
neighboring surface areas, which is a nonlinear coupling reminiscent
of the Josephson coupling in superconductor junctions
\cite{golestan97}. We focus on the case of sinusoidal corrugations
with a single wavelength, and show that it is possible to get a
rectified motion of the pinion. This is a manifestation of a
spontaneous symmetry breaking in the system, which is due the
inherent nonlinear structure of the dynamics. The average pinion
velocity of the rectified motion takes on discrete values and is set
by the wavelength of the corrugation $\lambda$ and the frequency of
vibration. We have also observed that the system shows a chaotic
behavior in a large part of the parameter space. We have studied the
effect of external load, and found an upper limit for the load it
can work against which is set by the amplitude of the lateral
Casimir force. We note that a similar study to ours has been
recently performed by Emig, in which he considers vibrations in the
normal distance between otherwise stationary corrugated surfaces
\cite{emig}.

\begin{figure}
\includegraphics[width=0.9\columnwidth]{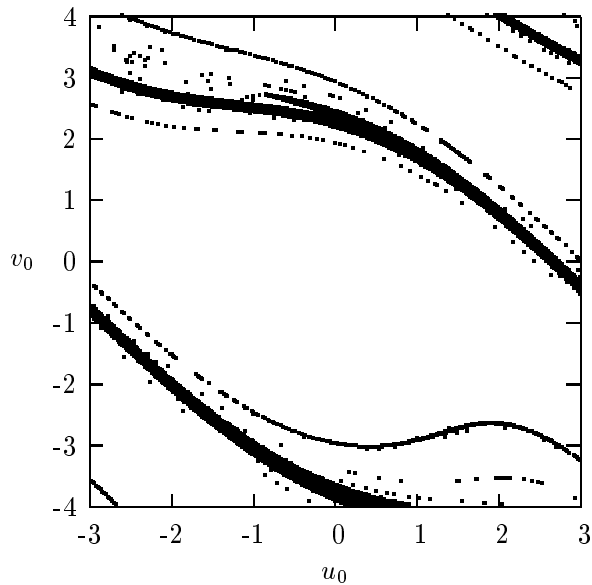}
\caption{Basin of attraction for $y_s=1.4$, $\omega_s=2/3$,
$\phi_s=0 $, $\epsilon=0.5$, $m=n=1$, and $w=0.185$. These initial
conditions lead to $V_{\rm P} >0$.} \label{fig:basin1}
\end{figure}

If two corrugated surfaces are placed parallel to each other, they
experience a lateral Casimir force that tends to displace them so
that their ``cogs'' will face each other to minimize the Casimir
energy. The force exerted on the lower plate reads \cite{golestan97}
\begin{equation}
F_{\rm lateral}= -F \sin\left[\frac{2 \pi}{\lambda}
(x-y)\right],\label{F-lat}
\end{equation}
where $x-y$ is the lateral relative displacement, $\lambda$ is the
corrugation wavelength of the two surfaces, and $F$ is the amplitude
(see below) \cite{Mohideen,Emig-etal-2001}. In the setup shown in
Fig. \ref{fig:gear}, we are interested in the dynamics of the pinion
of radius $R$ which is subject to a net torque $R F_{\rm lateral}$
due to the lateral Casimir force, in addition to other forces such
as an external load $W$. We can write the equation of motion for the
coordinate $x=R \theta$ ($\theta$ representing the rotation
dynamics) as
\begin{equation}
\frac{I}{R} \frac{d^2 x}{d t^2}=-R F \sin\left[\frac{2 \pi}{\lambda}
(x-y)\right]-\frac{\zeta}{R}\frac{d x}{d t}-r W, \label{master1}
\end{equation}
where $I$ is the moment of inertia about its major axis, and $\zeta$
is the rotational friction coefficient, and $r$ is the torque arm
for the external load.

\begin{figure}
\includegraphics[width=0.9\columnwidth]{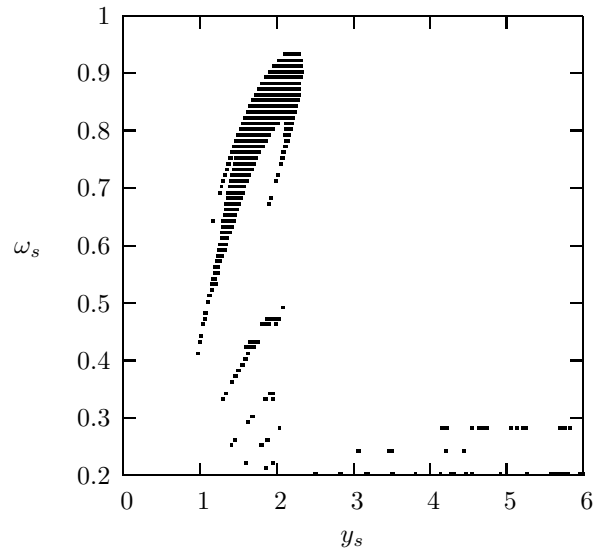}
\caption{Values of $y_s$ and $\omega_s$ which lead to a net rotation
of the pinion described by the velocity given in Eq. (\ref{Vp-vib})
with $m=n=1$. Here $x_0=0$, $\dot{x}_0=0$, $\epsilon=0.5$, and
$w=0.1$.} \label{fig:basin2}
\end{figure}

We can study the above nonlinear equation in the phase plane
$\left(u, v=\dot{u} \right)$ where $u \equiv 2 \pi (x-y)/\lambda$,
and the rescaled time $\tau$ is measured in units of $T=\sqrt{I
\lambda /(2 \pi F R^2)}$, namely $t=T \tau$. We consider harmonic
motion of the plate, i.e.
\begin{equation}
y=y_p \cos(\omega_p t + \phi_p),
\end{equation}
and rewrite Eq. (\ref{master1}) as
\begin{equation}
\ddot{u}=-\sin u -\epsilon \dot{u}-w+ y_s\cos(\omega_s \tau+
\phi_s), \label{master3}
\end{equation}
where $\epsilon= T \zeta /I $, $w=r W/(R F)$, $ \omega_s= \omega_p
T $, and
\begin{eqnarray}
y_s&=&\left(\frac{y_p \omega_p I}{F R^2}\right)
\left[\omega_p^2+\frac{\zeta^2}{I^2} \right]^{1/2},\label{ys} \\
\phi_s&=&\phi_p -\tan^{-1}\left(\frac{\zeta}{I
\omega_p}\right).\label{phis}
\end{eqnarray}
The dynamics of the system is described by Eq. (\ref{master3}) which
depends on five (dimensionless) parameters: $\epsilon$ that measures
the relative importance of friction to inertia, $w$ that is the
external load, and the amplitude $y_s$, the frequency $\omega_s$,
and the initial phase mismatch $\phi_s$ of the oscillatory input
motion.

Instead of embarking on a full study of the phase diagram of this
system, we focus on the specific cases of average rectified motion.
Following Refs. \cite{bel,Gwinn}, we seek for pairs of period
steady-state solutions $(u^*,v^*=\dot{u}^*)$ such that
\begin{eqnarray}
& & u^*(\tau+2 \pi n /\omega_s)=u^*(\tau)\pm 2 \pi m,  \label{u*} \\
& & v^*(\tau+2 \pi n /\omega_s)= v^*(\tau),\label{v*}
\end{eqnarray}
where $m$ and $n$ are integer numbers. We also focus on the upward
motion of the load or positive average velocities. The average
pinion velocity in this case can be deduced from $\overline{v^*}=m
\omega_s/n $, namely
\begin{equation}
V_{\rm P}=\frac{m}{n} \; \left(\frac{\lambda \omega_p}{2
\pi}\right).\label{Vp-vib}
\end{equation}
The steady-state solutions of Eqs. (\ref{u*}) and (\ref{v*}) are
only valid for specific values of the initial conditions $u_0$ and
$v_0$, and therefore the initial values must be tuned to ensure
$V_{\rm P}>0$. In Fig. \ref{fig:basin1}, we have shown the basin of
attraction for solutions with $V_{\rm P}>0$, using $y_s=1.4$,
$\omega_s=2/3$, $\phi_s=0$, $\epsilon=0.5$, $m=n=1$, and $w=0.185$.
The values of the initial conditions $u_0$ and $v_0$ have been
changed by increments of $0.02$ across the phase plane.

Equation (\ref{Vp-vib}) shows that the pinion velocity for the
rectified motion could only acquire distinct values set by the two
integers $m$ and $n$, and the overall scale is set by the vibration
frequency of the rack $\omega_p$ and the wavelength of the
corrugations $\lambda$. It is interesting that the driving
parameters can be tuned to set the integers $m$ and $n$. Figure
\ref{fig:basin2} shows the parameters $y_s$ and $\omega_s$ that lead
to $m=n=1$, using $x_0=0$, $\dot{x}_0=0$, $\epsilon=0.5$, and
$w=0.1$. In these calculations the steps of $y_s $ and $\omega_s$
are taken to be $0.01$. Throughout the numerical calculations, we
noticed that a large part of the parameter space leads to chaotic
behavior. These states were unwanted for our purposes, and have been
systematically avoided by checking the negativeness of the largest
Lyapunov exponent of the system \cite{Ar}.

The external load against which the pinion is doing work puts a
significant impediment on the motion and one wonders if there is a
critical value for the load that this device can tolerate, similar
to the problem of uniformly moving racks \cite{AFR1}. To obtain this
limiting value, we can average the equation of motion over the time
interval $2 \pi n /\omega_s$, which yields
$-w-\epsilon\overline{v^*}=\overline{\sin u^*}$. This result
necessitates
\begin{equation}
W< \frac{R F}{r} \left(1-\epsilon \omega_s \frac{m}{n}\right).
\end{equation}
However, this is only an upper bound for the critical load and one
needs to do more to find stronger criteria. We can make an attempt
towards this end by using the anzats $u^*\approx u_0+
\overline{v^*}\tau+ \Delta n/\omega_s \sin( \omega_s \tau/n)$ and
the identity $\exp(i z \cos \beta)= \sum_{k= -\infty}^{\infty} i ^k
J_k(z) \exp(i k \beta)$ to expand the sinus in terms of Bessel
functions. Using these, a stronger condition of
\begin{equation}
W<\frac{R F}{r} \left[\left|J_n\left(\frac{n \Delta}{\omega_s}
\right) \right|-\epsilon\omega_s \;\frac{m}{n} \right],\label{Wc}
\end{equation}
can be obtained. Our numerical calculations with $y_s=1.4$,
$\omega_s=2/3$, $\phi_s=0$, $\epsilon=0.5$, and $m=n=1$ show that
the upward motion of the load ceases at $w=0.192$, consistent with
our analytical estimate of Eq. (\ref{Wc}), which yields $w< 0.249$
using $J_1(\Delta/ \omega_s )< 0.582$.

\begin{figure}[t]
\includegraphics[width=0.92\columnwidth]{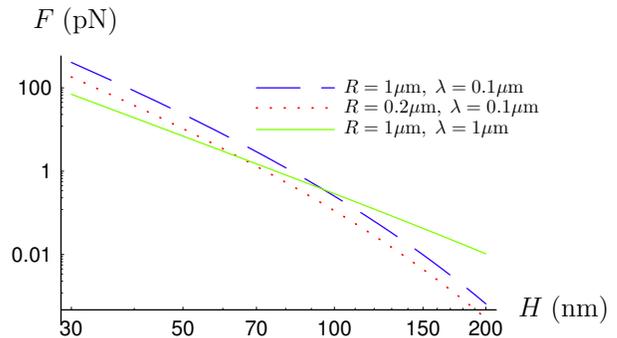}
\caption{(Color online) Amplitude of the lateral Casimir force as a
function of the gap size for perfect metallic boundaries,
corresponding to $a_1=a_2=10$ nm and $L=10 \mu$m, for different
values of radius and corrugation wavelength. Note that for the case
of $R=0.2 \mu$m (dotted curve), the larger separations in the plot
are approaching the limit of validity of the PFA approximation.}
\label{fig:F}
\end{figure}

The value of the critical load is essentially set by the amplitude
of the lateral Casimir force $F$. We can estimate this quantity for
the specific geometry of our rack and cylindrical pinion by using
the corresponding Casimir interaction between two parallel
corrugated plates and the Proximity Force Approximation (PFA)
\cite{prox}. For a cylinder of radius $R$ located at a (nearest)
distance $H$ from a plate, it has been shown recently that this
approximation is reasonably accurate for $H < R$
\cite{Kardar-cyl,bordag}. Using PFA for the lateral Casimir force
between a pinion of length $L$ and corrugation amplitude $a_1$ and a
rack of corrugation amplitude $a_2$, we find
\begin{equation}
F=\frac{\pi \sqrt{2} \hbar c a_1 a_2 L R^{1/2}}{\lambda H^{9/2}}
\;\int_1^\infty \frac{d s}{s^5
\sqrt{s-1}}J\left(\frac{H}{\lambda}s\right),\label{F-res-1}
\end{equation}
provided with $a_1, a_2 \ll H$. In this equation $J(u)$ is the
coupling function presented in Ref. \cite{Emig-etal-2001}. This
result shows a strong power law behavior at small values of $H$
followed by an exponential decay at large $H$ with the length scale
being set by $\lambda$. Figure \ref{fig:F} shows the value of the
amplitude $F$ as a function of the shortest distance between the
rack and the pinion, for various values of the wavelength and
cylinder radius. One can see that the amplitude and thus the maximum
load of the system strongly depends on the separation. While typical
values for $F$ is in the pN range for the parameter values we chose,
its linear dependence on the length of the pinion $L$ can be used to
strengthen the hold of the pinion and make it capable of enduring
higher load.

The fact that the system can spontaneously break the symmetry and
choose one direction is fascinating and far from trivial. Note that
the traditional theory of ratchets always relies on some kind of
asymmetry in the structure of the system (such as the two unequal
halves of an oscillating saw-tooth potential \cite{ratchet}), to
break the symmetry and guide the rectification process. This means
that the nonlinear structure of the dynamics of this system,
together with the presence of the inertial term, has made the
spontaneous symmetry breaking possible.

In conclusion, we have studied the dynamics of a set-up of rack and
pinion that are coupled by the lateral Casimir force, and shown that
the coupling can rectify an oscillatory motion in certain areas of
the parameter space of the system. This study could shed some light
on the novel possibilities that could be available at small scales
in systems that are powered by the Casimir force. This method of
force transduction could in principle help solve the wear problem in
nano-scale mechanics, which makes such studies worthwhile.


This work was supported by EPSRC under Grant EP/E024076/1 (R.G.).

\end{document}